\documentclass[12pt]{iopart}

\usepackage{graphicx}
\newcommand{\ii}{\'\i}
\newcommand{\M}[1]{\textrm{\textbf{#1}}}

\begin{document}

\title[]{Tracing the equation of state and the density of 
cosmological constant along z}

\author[]{Cristina Espa\~{n}a-Bonet and Pilar Ruiz-Lapuente}

\address{Departament of Astronomy, University of Barcelona, \\
CER en Astrof\ii sica, F\ii sica de Part\ii cules i Cosmologia i \\
Institut de Ciencies del Cosmos, Universitat de Barcelona ICCUB \\
Diagonal 647, Barcelona E--08028, Spain \\
}
\ead{cespana@am.ub.es, pilar@am.ub.es}

\begin{abstract}
We investigate the equation of state $w(z)$
in a non--parametric form using the latest compilations of distance luminosity
from SNe Ia at high z. We combine the inverse problem approach with a 
Monte Carlo to scan the space of priors. 
On the light of the latest high redshift
supernova data sets, we reconstruct 
$w(z)$. A comparison
between a sample including the latest results at z $>$ 1 and a sample without 
those results show the improvement achieved by observations of very high z 
supernovae. We present the prospects to measure  the 
variation of dark energy density along z by this method. 

\end{abstract}


\section{Introduction}
\label{intro}

The acceleration of the rate of expansion of the Universe,  
first discovered through supernovae \cite{Riess98,p99} is being examined 
through more extended samples of those distance indicators.
In the next decade, large samples of supernovae and an increased 
coverage in z of the measurements of baryon acoustic oscillations (BAO)
would allow to obtain the expansion rate $H(z)$ and the equation of state
$w(z)$ \cite{eisenstein,wang07}. 
Complementary information on the matter density and global curvature 
provided by the CMB measurements from Planck 
 \cite{planck} and weak lensing surveys \cite{heavens} would enable 
to establish a better range of allowed regions in $w(z)$. 
SNe Ia together with
BAO measurements up to very high z can be combined  
to restrict the range of possibilities of the empirical behavior
of dark energy. By recovering the empirical behavior of $w(z)$ one expects
to test whether the acceleration of the expansion of the Universe 
indicates modifications of gravity beyond GR, whether it is due to 
the presence of vacuum energy or it is associated with a light scalar
field  (\cite{padma, sahni, copeland, ruiz-lapuente, nobbenhuis,
 polarski06, dejan, padma07, roy07,elizalde07}).

The recovery of the function
 $w(z)$ from a given sample of data has been attempted
proposing fitting functions or expansion series of $w(z)$ along z 
in ways to accomodate a wide range of dark energy candidates. 
There has been some debate on the effect  that choosing particular 
models for those functions or truncating the expansion series in z
might have in deriving possible evolution 
\cite{bassett2,bruce, riess07}.

Here, we use an approach to obtain $w(z)$ without imposing
any constraints on the form of the function. This is obtained through
a generalized nonlinear inverse approach.
The inverse approach formulated by Bakus and Gilbert \cite{backus} 
has been widely used in geophysics and solar structure physics. 
In this approach,  
the mere fact that  the continuous functional has to be derived from a 
discrete number of data implies the non--uniqueness of the answer.
It has also been  shown that, 
even if the data were dense and with
no uncertainty, there would be more than one solution to many specific
inverse problems such as the determination of the density structure of 
the earth from the data on the local gravitational field  
at its surface, and others. 
This lack of uniqueness comes from the way in which the different
equations reflect in the observables used. The problem 
of the determination of dark energy faces such degeneracy. 
In the luminosity distance along z from supernovae  
and other cosmic distance indicators, $w(z)$ enters in an  
integral form, which limits the possibility to access to $w(z)$. 
In earlier examinations of the degeneracy in $w(z)$ obtained through
cosmic distance indicators, a range of solutions giving the same 
luminosity distance along z were pointed out \cite{maor1}. 
As more data would constrain $w(z)$ at various redshifts the
reconstruction should become more successful. Here, we examine this
using the latest compilations by Wood--Vasey et al (2007) \cite{essence}
and Davis al (2007) \cite{davis07} which use supernovae gathered by
many collaborations and add new ones from ESSENCE, SNLS
and the Higher--Z Team.

To compare with a significant body of work which analyses the data 
using the expansion to first order  $w(z)= w_0 + w_a z/(1+z)$ 
\cite{linder,polarski}, we will
also formulate  this approach for the case of determination of
discrete parameters. We examine from current data 
the possibility of determining at present the values of $w(z)$ 
and its first derivative and compare with previous results. 
 It is known (see for
instance \cite{bruce}) that this approximation does not allow to recover
properly the value of $w$ at high z.
And it has been argued \cite{riess07} that this expansion
might bias $w_{a}$ towards faster evolution. 
 It is, however, a very useful
approach to restrict quintessence proposals. Given its widespread use, 
we include here the discrete case with the two parameters $w_{0}$ and
$w_{a}$. 

In the following section, we introduce the inverse approach and 
deduce the equations for both the continuous and the discrete case.
Then, we show the reconstruction of $w(z)$ with current SNeIa
data. Finally, we draw our conclusions.

\section{Inverse problem}
\label{sec_IP}

\subsection{Non--parametric non--linear inversion}
\label{continuous}

The inverse problem provides a powerful way to determine the values
of functional forms from a set of observables. This approach is useful
when the information along a certain coordinate, in our case information on 
$w(z)$, emerges in observables coupled with information at all other z as it
happens with the luminosity distance. 
Dark energy is here addressed using the
non--linear non--parametric inversion. Most frequently, 
when the parameters to be determined
are a set of discrete unknowns, the method used is a least
squares. But the continuous case, where functional forms are to be 
determined, requires a general inverse problem formulation.
The inverse method used here is a Bayesian 
approach to this generalization \cite{tarantola}.

We consider a flat universe with only two dominant constituents
(at present): cold matter and dark energy. Therefore we characterize 
the cosmological model by the density of matter, $\Omega_M$, and 
by the index $w(z)$ of the dark energy equation of state,
\begin{equation} 
w(z)=\frac{p(z)}{\rho(z)} \,.
\end{equation}
The vector of unknowns $\M{M}$ has then a discrete and a continuous 
component,
\begin{equation}
      \M{M}  = \left( \begin{array}{c}
                  \Omega_M        \\
                  w(z)            \\
                  \end{array}
           \right)\,.
\end{equation}

On the other hand, the observational data are SNe\,Ia magnitudes. 
We have a finite set of $N$ magnitudes, $m_i$, and consider the following 
theoretical equation, the magnitude-redshift relation in a flat
universe relating the unknowns to the observational data: 

\begin{equation}
m^{th}(z,\Omega_M,w(z)) = M +  5\log[D_L(z,\Omega_M,w(z))] - 5\log H_0 + 25 , 
\label{mag}
\end{equation}

\noindent where $D_L$ is the Hubble-free luminosity distance
\begin{equation}
D_L(z,\Omega_M,w(z)) = c(1+z) 
\int_{0}^{z}{\frac{H_0 dz^\prime}{H(z^\prime,\Omega_{M},w(z))}}
\label{dl}
\end{equation}
with

\begin{eqnarray}
H(z^\prime,\Omega_{M},w(z)) =
H_0~\sqrt{\Omega_{M}(1+z^\prime)^3 + \Omega_{X}(z^\prime)} 
\label{hubblep}
\end{eqnarray}

\begin{equation}
\Omega_{X}(z)=\Omega_{X}\exp{\left( 3 \int_{0}^{z}
{dz^{\prime}\frac{1+w(z^{\prime})}{1+z^{\prime}} }\right)}.
\label{densol}
\end{equation}

We redefine our data and convert the original SNe magnitudes to
dimensionless distance coordinates $y$: 

\begin{equation}
y_i\equiv\frac{\exp_{10}{((\mu_i+5\log H_0-25)/5)}}{c(1+z_i)} =
\int_{0}^{z_i}{\frac{dz^\prime}{\sqrt{\Omega_{M}(1+z^\prime)^3 +
\Omega_{X}(z^\prime) }}},
\label{f}
\end{equation}

\begin{equation}
\sigma_{y_i} = \frac{\ln 10}{5} y_i \sigma_{\mu_i} ,
\label{sf}
\end{equation}

\noindent where $\mu_i = m_i -M$ is the distance modulus.
With this definition, we deal directly with a function $y(\Omega_M,w(z))$, 
the only part which depends on the cosmological model.

After the corresponding transformations, the observables are now described by 
a vector of $N$ components, $y_i$, and by a covariance matrix, $\M{C}_y$.
 This method can handle correlated measurements, where non--diagonal elements 
$ C_{y_{i}y_{j}}$ are different from zero
(observations i and j being correlated). But, at present, those have not been
estimated for the composite samples of distance indicators. We would
then use: 

\begin{equation}
\label{coy}
      C_{y,ij} =  \sigma^2_{y_{i}}  \delta_{ij}
\end{equation}

\noindent
Similarly, the unknown vector of   
parameters is described by its {\it a priori} value, $\M{M}_0$, 
and the covariance matrix, $\M{C}_0$. The function describing $w(z)$ is 
expected to be smooth, and this leads to no null covariance between neighboring
points in z for $w(z)$. Thus, the covariance matrix $\M{C}_0$ has the form:

\begin{equation}
\label{com}
      \M{C}_0 = \left( \begin{array}{cc}
                     \sigma^2_{\Omega_M}   & 0        \\
                     0            	   & C_{w(z),w(z')} \\
            \end{array} \right)
\end{equation}

\noindent
where a  choice is made for the non--null covariance between  z and z$'$,
 $C_{w(z),w(z')}$. This choice is taken to be as general as possible.
It would define the smoothness required in the solution by setting the
correlation length between errors in z and z$'$ (this gives the length
scale in which the function can fluctuate between redshifts). The
 amplitude of the 
fluctuation of the function is given by the dispersion $\sigma_w$ at z.

Thus for a Gaussian choice,  $C_{w(z),w(z')}$  is described as:

\begin{equation}
C_{w(z),w(z')}=\sigma_w^2\exp\left(-\frac{(z-z')^2}{2\Delta_z^2}\right),
\label{covwgauss} 
\end{equation} 

\noindent
which means that the variance at z equals $\sigma_w^2$ and that the 
correlation length between errors is $\Delta_z$. 
Another possible choice for $C_{w(z),w(z')}$
is an exponential of the type: 

\begin{equation}
C_{w(z),w(z')}=\sigma_w^2\exp\left(-\frac{|z-z'|}{\Delta_z}\right),
\label{covwexp} 
\end{equation} 

\noindent
while no difference in the results is found by those different choices.

This is all the information we have beforehand, and with that 
we are interested in determining the best estimator ${\tilde{\M{M}}}$ 
for \M{M}.
The probabilistic approach we will use incorporates constraints from 
priors through the Bayes theorem, i.e, the {\it a posterori} 
probability density  $f_{post}(\M{M}/\M{D})$
for the vector \M{M} containing the unknown model parameters given the
observed data \M{D}, is linked to the likelihood 
function L and the prior density function for the parameter vector as: 

\begin{eqnarray}
 f_{post}(\M{M}/\M{D}) \, \, \alpha \, {\it L}(\M{D}/\M{M}) \, \, 
\cdot \, f_{prior}(\M{M}) 
\label{chi0}
\end{eqnarray} 

The theoretical model described by the operator $\M{y}^{th}$, which 
connects the model parameters \M{M} with the predicted data
$ \M{D}_{predicted} = \M{y}^{th}(\M{M})$, is to agree as closely as possible 
with the observed data $\M{y}$. Assuming that both the prior probability and 
the errors in the data are distributed as Gaussian functions, the
posterior distribution becomes:

\begin{eqnarray}
 f_{post}(\M{M}/\M{y}) \, \alpha \, \exp[\,- 
\frac{1}{2}~(\M{y}-\M{y}^{th}(\M{M}))^* ~ \M{C}_y^{-1} ~ 
(\M{y}-\M{y}^{th}(\M{M})) \nonumber   \\
- \frac{1}{2}~(\M{M}-\M{M}_0)^* ~ \M{C}_0^{-1} ~ (\M{M}-\M{M}_0)\,] 
\nonumber \\ 
\label{chi5}
\end{eqnarray} 

\noindent
where $^{*}$ stands for the adjoint operator. 
The best estimator for \M{M}, $\tilde{\M{M}}$, is the most probable value of
\M{M}, given the set of data $\M{y}$. The condition is reached by 
minimizing the misfit function:

\begin{eqnarray}
S \equiv \frac{1}{2}~(\M{y}-\M{y}^{th}(\M{M}))^* ~ \M{C}_y^{-1} ~ 
(\M{y}-\M{y}^{th}(\M{M})) +  
\nonumber \\
\frac{1}{2}~(\M{M}-\M{M}_0)^* ~ \M{C}_0^{-1} ~ (\M{M}-\M{M}_0),
\label{chi}
\end{eqnarray} 

\noindent
which is equivalent to maximize the Gaussian density of 
probability when data and parameters are treated in the same way. 
This Bayesian approach helps to regularize the
inversion.

Let us now define the operator $\M{G}$ represented by the matrix of partial 
derivatives of the dimensionless distance coordinate, which will simplify 
subsequent notation. Its kernel will be denoted by $g$ as defined in the next
equations.

\begin{equation}
\label{g}
\M{G} = \left( \begin{array}{cc}
                    \frac{\partial y_{1}^{th}}{\partial \Omega_M} & 
			\frac{\partial y_{1}^{th}}{\partial w(z)}      \\
                     \frac{\partial y_{2}^{th}}{\partial \Omega_M} & 
			\frac{\partial y_{2}^{th}}{\partial w(z)}     \\
			: & :	\\
		     \frac{\partial y_{N}^{th}}{\partial \Omega_M} &
			\frac{\partial y_{N}^{th}}{\partial w(z)}
                 \end{array}
           \right) 
\end{equation}

\noindent
with

\begin{eqnarray}
 \frac{\partial y_{i}^{th}}{\partial \Omega_M}&=& -\frac{1}{2}
\int_{0}^{z_i}\frac{(1+z')^3 dz'}{H^3(z')} 
 \equiv \int_{0}^{z_{i}} g_{\Omega_M}(z^\prime)dz^\prime,
\label{derom}
\end{eqnarray}

\begin{eqnarray}
\frac{\partial y_{i}^{th}}{\partial w(z)}&=& -\frac{1}{2}
\int_{0}^{z_i}\frac{3\Omega_X(z')\ln(1+z') dz'}{H^3(z')} 
\equiv  \int_{0}^{z_{i}} g_{w}(z^\prime)dz^\prime.
\label{derwz}
\end{eqnarray}

\noindent
As shown in Eq.\,\ref{mag} or equivalently Eq.\,\ref{f}, the inverse 
problem is nonlinear in the parameters, thus the solution 
is reached iteratively in a gradient based search. To minimize 
$S$ in Eq.\,\ref{chi}, one demands stationarity. For the nonlinear 
case the solution has to be implemented as an iterative procedure where
 \cite{tarantola}: 

\begin{eqnarray}
\label{solutionMit}
\tilde{\M{M}}_{[k+1]} = \M{M}_0 ~ + ~ \M{C}_0 ~ \M{G}^*_{[k]} ~ 
(\M{C}_y + \M{G}_{[k]} \M{C}_0 ~ \M{G}^*_{[k]})^{-1}
 \nonumber \\
(\,\M{y} ~ - ~ \M{y}^{th}(\tilde{\M{M}}_{[k]}) ~+~ \M{G}_{[k]} ~ 
(\tilde{\M{M}}_{[k]}-\M{M}_0)\,)
\end{eqnarray}

\noindent
Since we are
working in a Hilbert space with vectors containing functional forms,
the above operator products give rise to scalar 
products of the functions integrated over the domain of those 
functions. The expressions transform into having the products
rewritten in terms of the kernels of the operators
\cite{nercessian}.

\noindent
We will indicate the scalar product by `` $\cdot$ '' and it is 
defined as it can be seen from this example:

\begin{equation}
 C_w \cdot \frac{\partial y_{j}^{th}}{\partial w(z)} = 
\int_{0}^{z_j} C_w(z,z^\prime) g_w(z^\prime) dz^\prime
\end{equation}

The components of
the vector of unknowns $\tilde{\M{M}}$, which in our case will be 
both $\Omega_M$ and $w(z)$, are then obtained from:

\begin{eqnarray}
\tilde{M}_{[k+1]}(z) = M_0(z) ~ + ~ 
\sum_{i=1}^N W_{i[k]} \int_{0}^{z_i}C_{0}(z,z')g_{i[k]}(z')dz'\,,
\end{eqnarray}
where 

\begin{equation}
\label{w}
W_{i[k]} =  \sum_{j=1}^N \left(S_{[k]}^{-1}\right)_{i,j} V_{j[k]} 
\end{equation}

\begin{eqnarray}
\label{Sgen2}
\M{V}~~~ & = & \M{y} + ~\M{G}~ (\M{M}-\M{M}_0) - \M{y}^{th}(\M{M}) \nonumber \\
  V_{i[k]} & = & y_{i} + \int_{0}^{z_i} g_{i[k]}(z) \left( M_{[k]}(z)-M_0(z) 
\right)dz 
          - \, y_{i}^{th}(z_i,\Omega_M,w(z)) \\
\nonumber \\
  \M{S}~~~~~\, & = & \M{C}_y ~ + ~ \M{G}~ \M{C}_0 ~\M{G}^*  \nonumber \\
  S_{i,j[k]} & = & (C_y)_{i,j} + 
              \int_{0}^{z_j}\int_{0}^{z_i} g_{i[k]}(z) \,C_0(z,z')\, 
g_{j[k]}(z') \,dz\, dz' 
\end{eqnarray} 

In the case of the dark energy equation of state and the matter
density the expressions reduce to 

\begin{equation}
\Omega_{M[k+1]}= \Omega_{Mo} + \sigma_{\Omega_M}^2 \sum_{i=1}^N W_{i\,[k]}\,  
{\frac{\partial y_i^{th}}{\partial \Omega_M}}_{[k]}
\label{om}
\end{equation}

\begin{equation}
w_{[k+1]}(z) = w_o(z)+ \sum_{i=1}^N W_{i\,[k]} 
              \int_{0}^{z_i}C_{w}(z,z')g_{w[k]}(z')dz'
\label{wz}
\end{equation}

\noindent
where
 $C_{w}(z,z')\equiv C_{w(z),w(z')}(z,z')$, $ W_{i\,[k]}$
is given by the product (\ref{w}) with:

{\setlength\arraycolsep{2pt}  
\begin{eqnarray}
V_i & = & y_{i} + \frac{\partial y_{i}^{th}}{\partial \Omega_M} 
(\Omega_M-\Omega_{M_0}) + \frac{\partial y_{i}^{th}}{\partial w(z)} \cdot 
(w-w_o) 
 - \, y_{i}^{th}(z_i,\Omega_M,w(z)) \\
\nonumber \\
S_{i,j} & = &  \delta_{i,j} \sigma_i \sigma_j +
\frac{\partial y_{i}^{th}}{\partial \Omega_M} C_{\Omega_M} 
\frac{\partial y_{j}^{th}}{\partial \Omega_M} + 
        \frac{\partial y_{i}^{th}}{\partial w(z)} \cdot \left( C_w 
         \cdot \frac{\partial y_{j}^{th}}{\partial w(z)}  \right) 
\label{s}
\end{eqnarray}

To test the accuracy of the inversion we use the {\it a 
posteriori} covariance matrix. It can be shown (see 
\cite{tarantola2,nercessian})
that for the linear inverse problem with Gaussian {\it a priori} 
probability density function, the {\it a posteriori} probability density 
function is also Gaussian with mean Eq.\,\ref{solutionMit} and covariance 
Eq.\,\ref{cov}. Although its value is only exact in the linear 
case it is a good approximation here, since the luminosity distance is quite 
linear on the equation of state $w(z)$ at low redshift.

\begin{eqnarray}
\M{C}_{\tilde{\rm{M}}} &=& ( \,\M{G}^*\, \M{C}_y^{-1} \,\M{G} \,+\, 
\M{C}_0^{-1} \,)^{-1} \equiv 
\M{C}_0-\M{C}_0\,\M{G}^*\,\M{S}^{-1}\,\M{G}\,\M{C}_0 \nonumber \\
    &=&  (\,\M{I}-\M{C}_0\,\M{G}^*\,\M{S}^{-1}\,\M{G}\,)\,\M{C}_0
\label{cov}
\end{eqnarray}

\noindent
In an explicit form, the standard deviations from this covariance 
read

\begin{eqnarray}
\label{som}
\tilde{\sigma}_{\Omega_M} &=& \sqrt{C_{\tilde{\Omega}_M}} = 
 \sigma_{\Omega_M}
\sqrt{1- \sum_{i,j}\frac{\partial y_{i}^{th}}{\partial \Omega_M} 
(S^{-1})_{i,j} 
\frac{\partial y_{j}^{th}}{\partial \Omega_M}\sigma_{\Omega_M}^2}
\end{eqnarray}

\begin{eqnarray}
\label{swz}
\tilde{\sigma}_{w(z)}(z) &=& \sqrt{C_{\tilde{w}(z)}(z)} =  
\sqrt{ \sigma_{w(z)}^2 - \sum_{i,j} C_{w} \cdot 
\frac{\partial y_{i}^{th}}{\partial w(z)} (S^{-1})_{i,j} 
\frac{\partial y_{j}^{th}}{\partial w(z)} \cdot C_{w}} 
\end{eqnarray}

\noindent
where the symbols with tilde are the {\it a posteriori} values,
whereas the symbols without represent the {\it a priori} ones.
It must be stressed that the uncertainty in the final $w(z)$ does
depend on the {\it a priori} assumption of the uncertainty. In fact, 
the same $w(z)$ could depend on the prior. For avoiding a dependence 
of the result and its error on the prior we use 
Monte Carlo methods later in the analysis.

There are other parameters which help to interpret the results.
From the form of Eq.\,\ref{cov} we see that the operator 
$\M{C}_0\M{G}^*\M{S}^{-1}\M{G}$ is related to the obtained resolution. 
This is usually called the 
{\it resolving kernel} $K(z,z^\prime)$. The more this term resembles
the $\delta$-function the smaller the {\it a posteriori} covariance 
function is. In fact, in the linear case, the resolving kernel represents
how much the results of the inversion differ from the true model.
It equals to be the filter between the true model and its estimated
value \cite{backus, tarantola}. In any applied case, it is a
 low band pass filter 
which depends on the data available and the details requested from the model. 
 In a useful way, it can also be 
expressed in terms of the {\it a priori} and the {\it a posteriori} 
covariance matrices:
\begin{equation}
 \M{K}\, = \,\M{I}\, - \,\M{C}_{\tilde{\rm{M}}}~\M{C}_0^{-1}.
\label{K}
\end{equation}

\noindent 
This expression will be evaluated numerically to quantify the
resolution and information generated in the inversion.

\subsection{Discrete parameters}
\label{discret}

In the previous section we have obtained the results for a set  
of a continuous function and a discrete parameter, 
but we can also consider the case 
of various discrete parameters. It was pointed out that a succesful
parameterization for modeling a large variety of dark energy models
is obtained by considering $w(z)$ expanded around the scale factor
$a$. The earlier parameterization to first order in z given 
by $w(z)= w_0 + w' z$ 
proved unphysical for the CMB data and a poor approach to SN data at z
$\sim$ 1. 
  For the case of moderate evolution in the equation of state, 
the most simple
(two--parameter) description of $w(z)$ so far proposed is
\cite{polarski, linder}: 

\begin{equation}
w(z)= w_0 + w_a (1 -a)
\label{EOS0}
\end{equation}

\noindent
where the scale factor $a = (1+z)^{-1}$ and $w(z)$ turns out to be: 

\begin{equation}
w(z)= w_0 + w_a \frac{z}{1+z}.
\label{EOS1}
\end{equation}

\noindent
 We use now this particular form for the 
function $w(z)$ commonly used to study the behaviour
of dark energy to solve iteratively for  $w_0$ and $w_a$:

\begin{equation}
w_{0[k+1]}= w_{0}^0 + \sigma_{w_0}^2 \sum_{i=1}^N W_{i\,[k]} 
\frac{\partial y_i^{th}}{\partial w_0}_{[k]}
\label{eqw0}
\end{equation}
\begin{equation}
w_{a[k+1]}= w_{a}^0 + \sigma_{w_a}^2 \sum_{i=1}^N W_{i\,[k]} 
{\frac{\partial y_i^{th}}{\partial w_a}}_{[k]}
\label{eqwa}
\end{equation}

\noindent 
where

\begin{equation}
 \frac{\partial y_i^{th}}{\partial w_0}= -\frac{1}{2}
\int_{0}^{z_{i}}\frac{3\Omega_X(z^\prime)\ln(1+z^\prime) dz'}{H^3(z')},
\end{equation}
\begin{equation}
 \frac{\partial y_i^{th}}{\partial w_a}= -\frac{1}{2}
\int_{0}^{z_{i}}\frac{3\Omega_X(z^\prime)[\ln(1+z^\prime)-\frac{z^\prime}
{1+z^\prime}] dz'}{H^3(z')}.
\end{equation}

The {\it a posteriori} variance is for these parameters:

\begin{equation}
\tilde{\sigma}_{w_0} = \sqrt{C_{\tilde{w}_0}} = \sigma_{w_0}
\sqrt{1- \sum_{i,j}\frac{\partial y_{i}^{th}}{\partial w_0} (S^{-1})_{i,j} 
\frac{\partial y_{j}^{th}}{\partial w_0}\sigma_{w_0}^2}
\end{equation}
\begin{equation}
\tilde{\sigma}_{w_a} = \sqrt{C_{\tilde{w}_a}} = \sigma_{w_a}
\sqrt{1- \sum_{i,j}\frac{\partial y_{i}^{th}}{\partial w_a} (S^{-1})_{i,j} 
\frac{\partial y_{j}^{th}}{\partial w_a}\sigma_{w_a}^2}
\end{equation}

\noindent
The equations for $\Omega_M$ are those of section \ref{continuous}
(Eqs.\,\ref{om}, \,\ref{derom} and \,\ref{som}).

\section{Determination of $\lowercase{w(z)}$}

 In the following, we determine $w(z)$ using the inverse approach described 
above and the SNeIa by Davis et al. \cite{davis07} which contains
samples of high--z supernovae 
by various collaborations (the Supernova Cosmology Project, the 
High--Z Team, SNLS, the Higher--z team and ESSENCE). 
  
Explicitly, one obtains the value of $\Omega_M$ and $w$ 
at a given redshift with the equations of section \ref{continuous}
(Eqs.\,\ref{om}, \ref{wz}, \ref{som} and \ref{swz}). In all the cases, it is
assumed a flat universe, where the equations have been deduced, and a good
knowledge for the density of matter: $\Omega_M= 0.27 \pm 0.03$.

The {\it a priori} model as we will show is arbitrary for $w(z)$, but 
determines where the solution is searched. To avoid the possibility 
of having converged to a secondary minimum
just because  the prior is too far from the absolute one, we have also 
carried out a Monte Carlo (MC) exploration of the {\it a priori} model 
space. A flat distribution of 1000 models between $-3<w(z)<1$ has been 
randomly generated, and then the iterative process of inversion has been 
applied for all of them. We used for this inversion 1000 data sets 
obtained by bootstrap resampling of the original one. The MC exploration and
the bootstrap approach   
allows to obtain the variance and deviation at every z in the non--parametric 
 function $w(z)$.
The resulting $1\sigma$ is obtained in this way. 

 Our results show with a Monte Carlo exploration with 
priors ranging from $w(z)=-3$ to $w(z)=+1$ that the obtained solution is 
stable.

\begin{figure}
  \begin{center}
    \scalebox{0.55}{\includegraphics{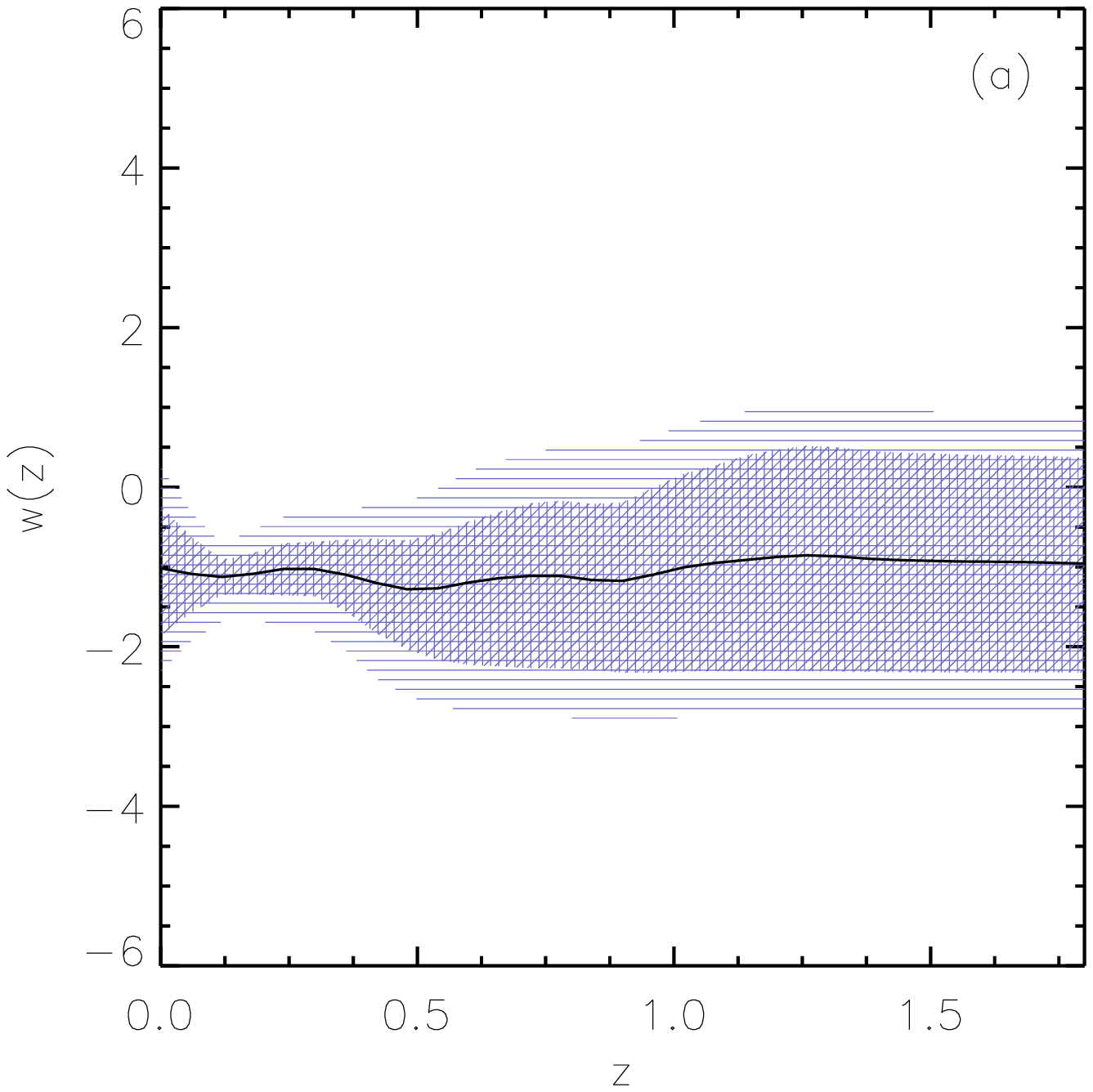}} 
    \scalebox{0.55}{\includegraphics{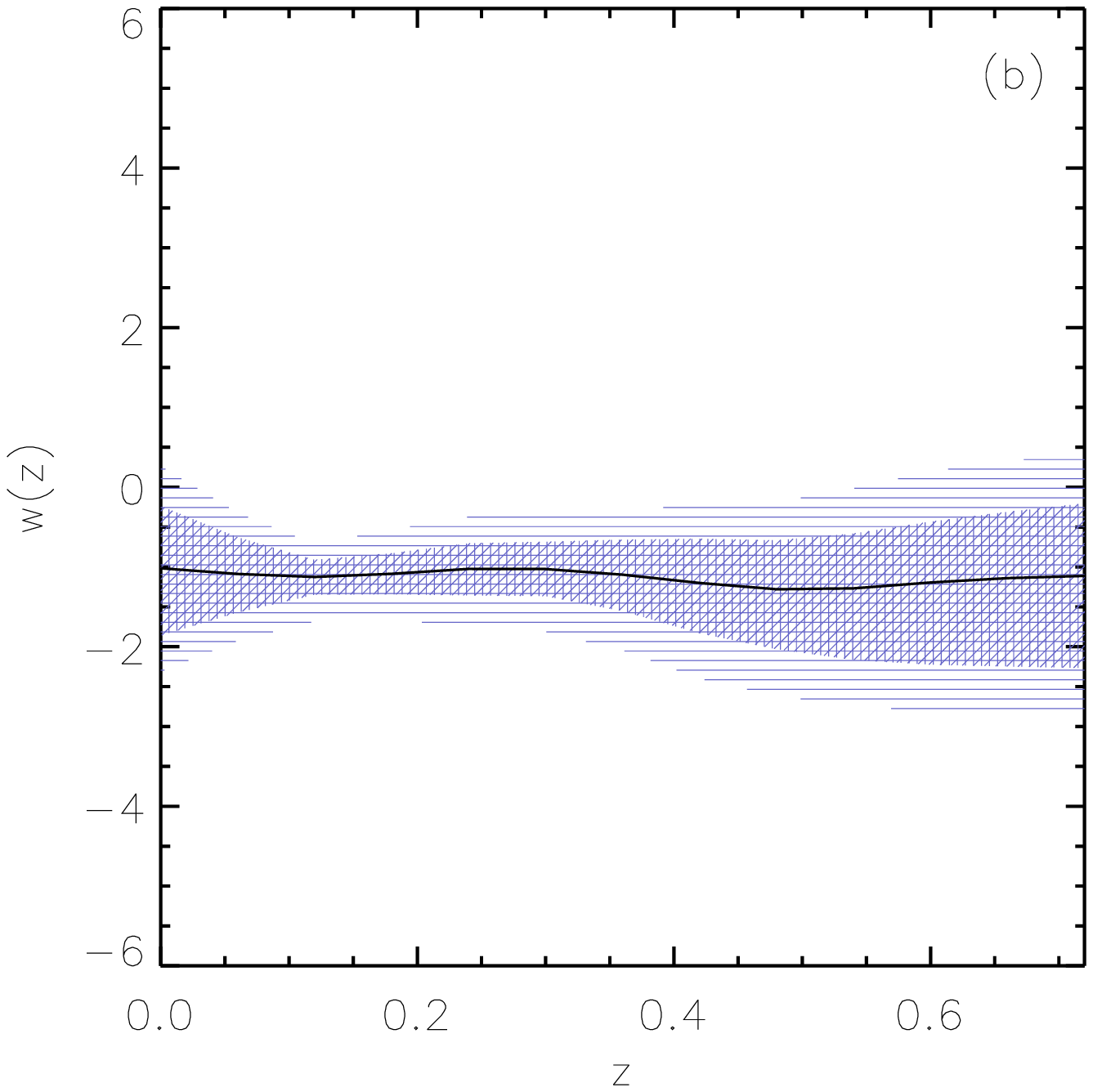}} 
  \end{center}

\caption{(a) Inversion of 1000 data sets generated by bootstrap resampling of 
the latest high redshift compilation in \cite{davis07}. Initial priors on the 
equation of state are randomly distributed between $-3<w(z)<1$. The solid black line represents the mean of the 1000 inversions and the filled regions are the 
intervals where $1\sigma$ and $2\sigma$ of the results lie. (b) The same as 
(a) for the redshift interval of best reliability.
 The panel shows $w(z)$ (solid line) and the 
$1\sigma$  (dashed shadow) and $2\sigma$ (pale dashed 
shadow) confidence intervals.}
\label{fig:wzMCboot}
\end{figure}

The solution using the 192 supernovae (60 from ESSENCE, 57 from SNLS and 
45 nearby ones and 30 by Riess et al. 2006) is 
shown in Figure\,\ref{fig:wzMCboot}. The calculation uses redshift intervals 
of $\delta$z $=$ 0.06. Various functional forms 
for the covariance have been tried giving the same results (covariance as 
in Eq.\,\ref{covwexp} or the Gaussian Eq.\,\ref{covwgauss} give similar 
results). This calculation does not restrict   
fast evolving  $w(z)$ as the correlation 
length is kept to allow fast changes of slope
($\Delta_z=0.08$). In a situation where the data
 are scarce with very wide priors 
in the function
$w(z)$, the solution can iterate between saddle points and local
minima as few data do provide a landscape with no strong minima.
However, in our case, the sample available for $w(z)$ 
is large enough to allow to find the solution.

As it has been seen in Figure 1, at low and intermediate redshift the data 
delimit a narrow band of possible solutions. At low and intermediate redshift,
where the inversion is reliable, there is no 
sign of any significant evolution. The cosmological constant is always 
within $1\sigma$ intervals.

 In Figure 1 the black solid line indicates 
the evolution of the equation of state, whereas the shadowed regions
 represents 
the 1$\sigma$ and 2$\sigma$ intervals. In Figure 2 the resolving
kernels at z $=0,0.24,0.48,0.84$ and $1.20$ are shown. It can be seen that
only the prior is recovered at high redshift, where there is no information 
coming from data and the resolving kernels are almost flat. While Figure 1
show the results for the inversion using the \cite{davis07} sample, very
similar results are found if using the sample from \cite{essence}. 
The resolving kernels (Figure 2) indicate that the function
is generally not well resolved at individual redshifts, well beyond 
the redshift range z $\sim 0.5$. At high redshift, z $=1.2$ for example, we 
observe a very wide and extremely flat $K(z,1.2)$, meaning that this redshift 
is not resolved at all by the data. The reliability of the inversion peaks in 
the range of z $\sim$  0.2--0.5 where the information is maximum. 
This is also found in other analyses, though most of the
$w(z)$ reconstructions \cite{trotta07,nesseris,riess07,periva}
have been done prior to the availability of these combined samples.

\begin{figure}
  \begin{center}
    \scalebox{0.55}{\includegraphics{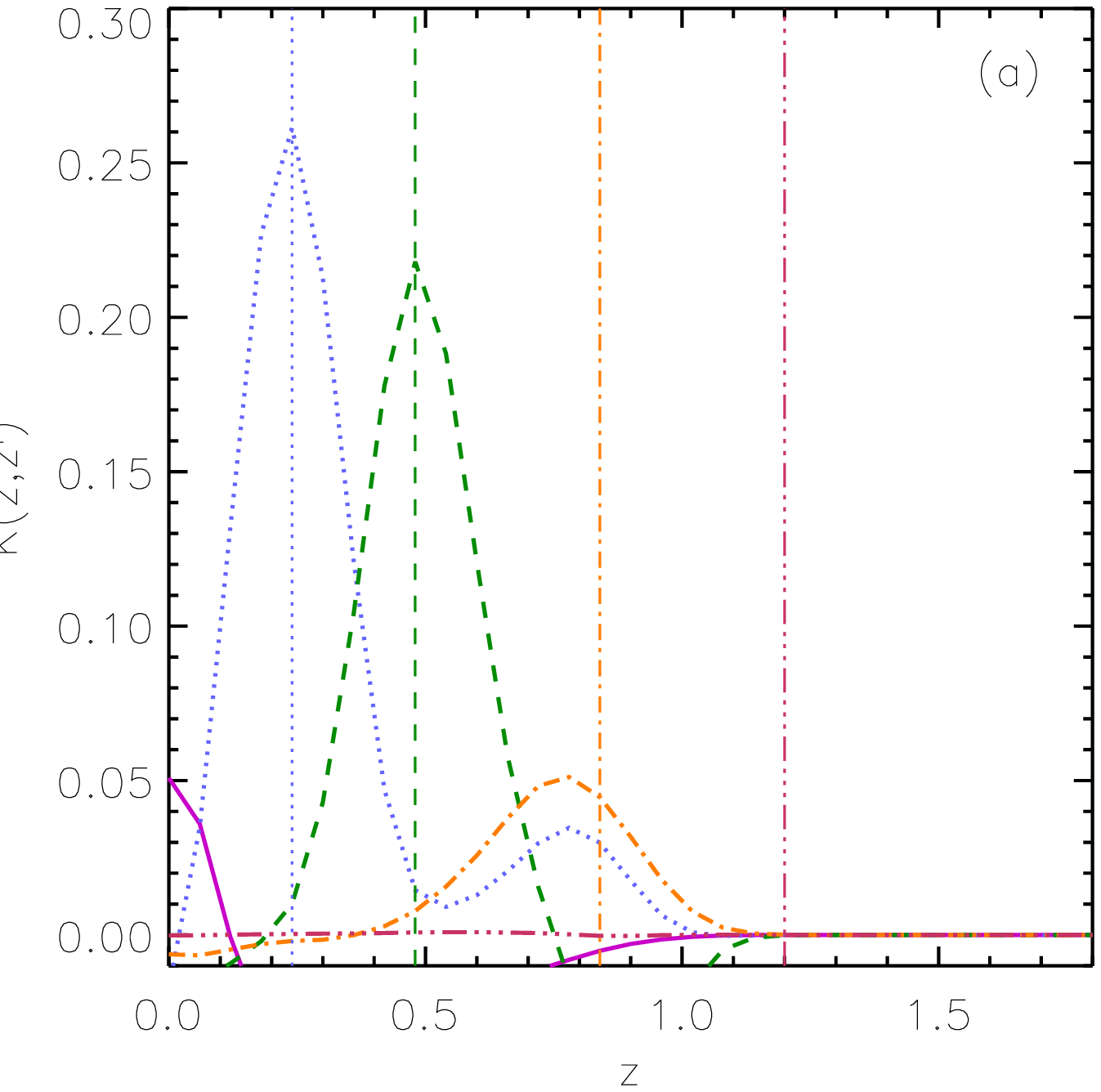}} \\
    \scalebox{0.55}{\includegraphics{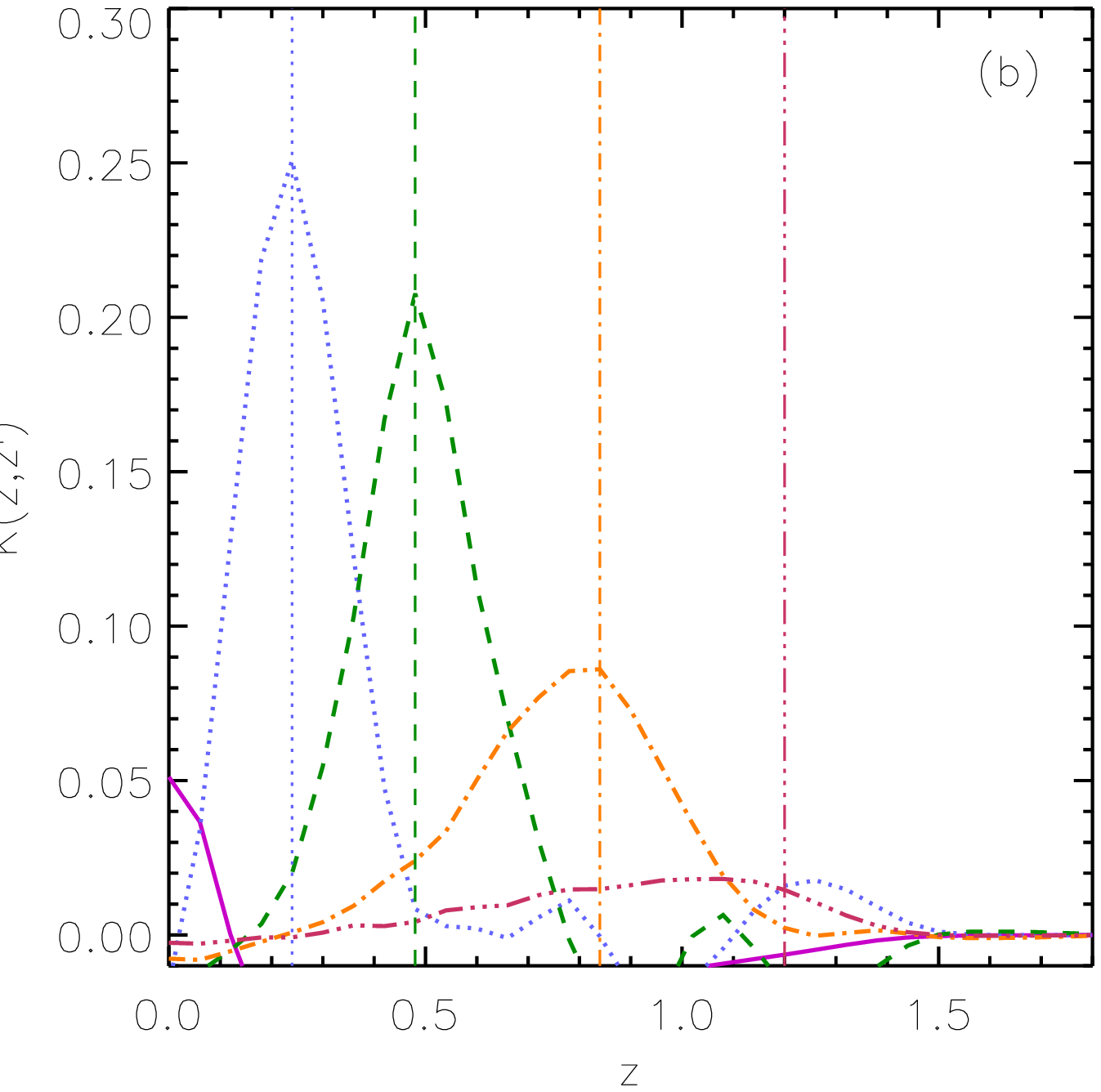}} 
  \end{center}

\caption{{\it Top panel.} Reconstruction of $w(z)$ using the 162 SNe
in Ref \cite{essence}.
The different 
resolving kernels at z $=0,0.24,0.48,0.84,1.20$ are shown.
 The resolving kernels at high z show that there is no information to 
conclude on the evolution of the equation of state. there. {\it Bottom panel.}
The same as before but for the sample in Ref.~\cite{davis07}. That allows 
to see the effect of the HST supernovae on the evolution at intermediate 
redshift.}
\label{fig:wz192}
\end{figure}

Finally, one can particularize the continuos form of $w(z)$ to a 
parameterization and compare the results with other analyses. Using the 
equations from Section~\ref{discret} for $\Omega_M$, $w_0$ and $w_a$ with 
priors $\Omega_M=0.27\pm0.03$, $w_0=-1\pm10$ and $w_a=0\pm10$ we 
obtained a positive evolution of the equation of state for the full data 
set: $w_0=-1.1\pm 0.3$, $w_a=0.6 \pm 1.6$ (when excluding the highest z
HST supernovae  $w_0=-0.7 \pm 0.4$, $w_a=-2 \pm 2$). 
This result agrees with the
continuous evolutions seen before, as it is expected for a smooth and globally
monotonous behaviour. For a constant equation of state, both data 
sets favour a cosmological constant within its $1\sigma$ intervals with a 
value of $w_0=-1.01\pm0.13$ in the first case and $w_0=-1.08\pm0.14$ 
in the second one.

\section{A running cosmological constant as an inverse problem}

We now use the
power of the inverse method to detect
an evolution of the cosmological constant. 
We interpret the unknown function as a running cosmological constant, 
but it can be interpreted as well as a general function
of the dark energy density.

\smallskip

\noindent
A running lambda can occur in
various different scenarios (see for instance \cite{nobbenhuis,espanabonet} 
and references therein). 
We want to determine
a general function $\Delta\Omega_{\Lambda}(z)$ such that

\begin{equation}
\Omega_{\Lambda}(z)= \Omega_{\Lambda}^0 + \Delta\Omega_{\Lambda}(z)
\label{sum}
\end{equation}

\noindent
where $\Omega_{\Lambda}^0$ is the value of $\Omega_{\Lambda}(z)$ for z $=$ 0. 
We use equations Eq. \,\ref{dl} and \ref{hubblep} to obtain the $\M{G}$
matrix.
We calculate the
form of the best $\Omega_{\Lambda}(z)$ by computing $\partial y^{th}/
\partial\Delta\Omega_{\Lambda}(z)$ in iterations until numerical convergence
is obtained.

We tested various degrees of prior 
knowledge on $\Omega_{M}$.  
We display here the situation where we have a very precise 
knowledge of $\Omega_{M}$ as it is expected after the Planck
mission. If we knew well the matter density, we would
see the evolution of $\Omega_{\Lambda}(z)$, as in Figure 3.

The procedure for the inversion is the same as for the barotropic index of
the equation of state.
The uncertainty on the prior on 
$\Delta\Omega_{\Lambda}(z)$ is set to
$\sigma(z) = 0.1$. 
The Monte Carlo exploration of the $\Omega_{\Lambda}(z)$-space is made
in the range $-0.2<\Delta\Omega_{\Lambda}(z)<0.2$. Since we are using
the same data sets as in previous sections, we expect the same resolution.
Therefore we use the same grid resolution.

For the two samples, {\rm VW07}  and {\rm D07}, the result of the
inversion is quite similar. A constant density cannot be discarded.
The reconstruction can not provide further information beyond 
redshift 0.6, but gives the behavior in 
the range 0 to 0.6.  If we relax the prior knowledge of $\Omega_{M}$
to a range of 0.03 we find a similar mean behavior but with a wider
range of allowed values for $\Omega_{\Lambda}$. 
Figure 3 is shown for purposes of indicating the enormous value of 
the complementary information to be provided in the next future.

\begin{figure}
     \scalebox{0.52}
{\includegraphics{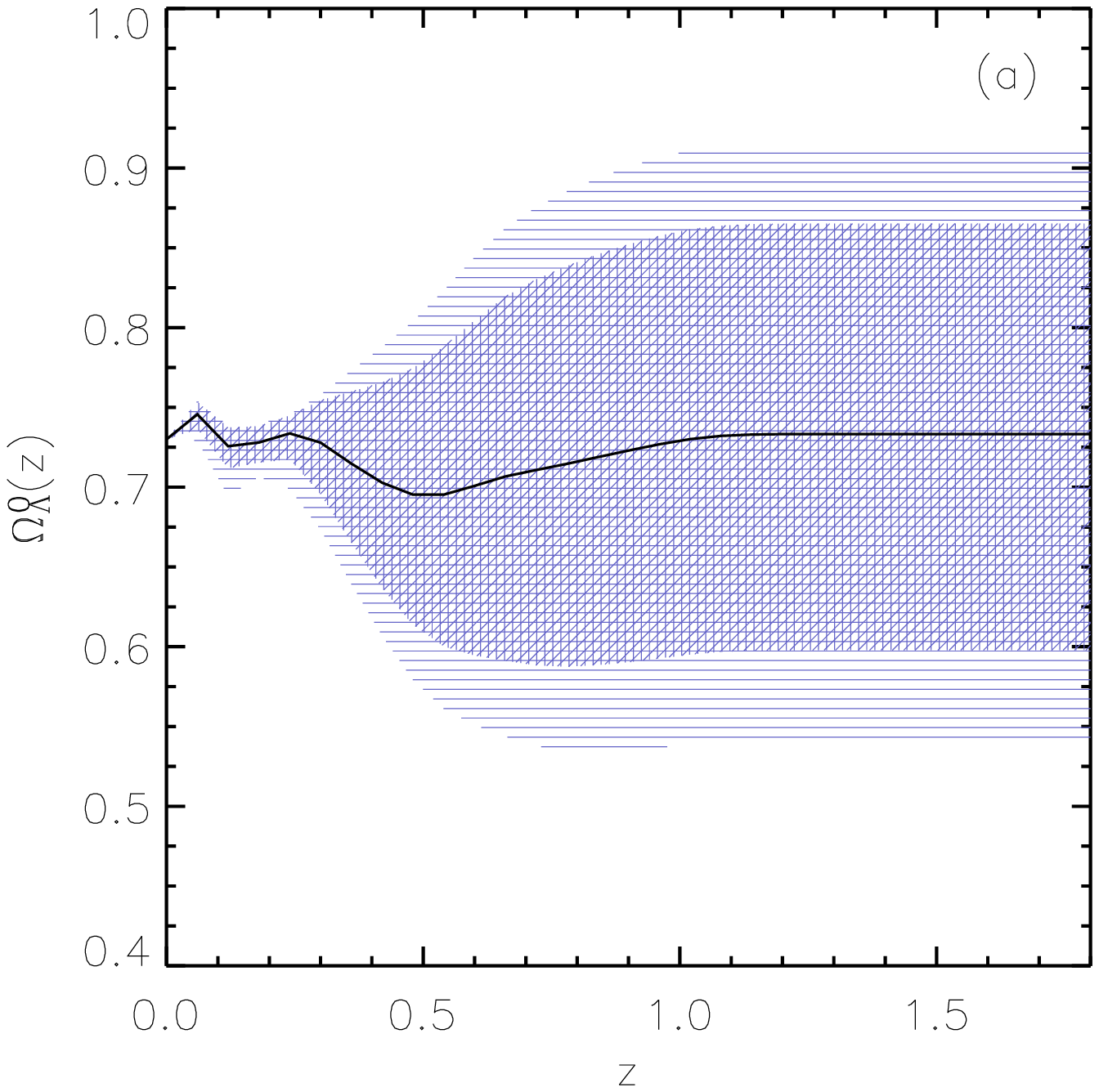}}
     \scalebox{0.52}
{\includegraphics{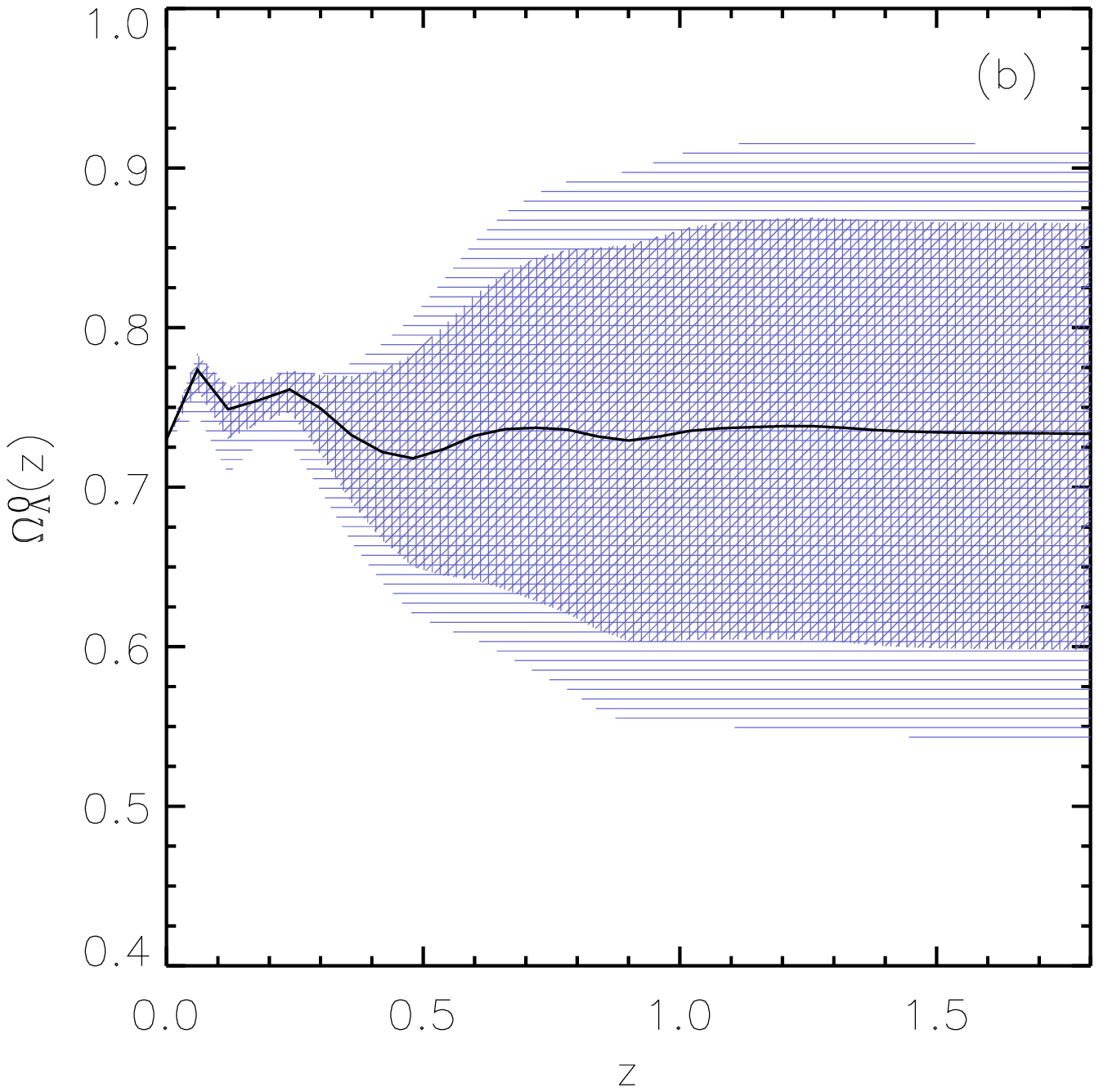}}

\caption{Reconstruction of $\Omega_{\Lambda}(z)$ for a flat Universe
 with $\Omega_{M}$ $=$ 0.27. The data compilation
in {\rm VW07} is used in the inversion shown in panel (a).
 The data compilation in {\rm D07} is used in the inversion shown in panel
 (b). The possibilities opened by precise measurements of $\Omega_{M}$  and
 further data at very high z are illustrated here.}
\label{fig:wzCCvar192}
\end{figure}

These reconstructions can be compared with the behavior of the cosmological
constant in the running scenarios. 
Up to now, the obtained smooth evolutions down to redshifts lower
than 1 are compatible with various scenarios for running cosmological
constant as no fast changes are expected in the past Gyrs. Scenarios of 
moderate evolution with $\nu$ $\sim$ 0.1 \cite{espanabonet} can be easily
discarded with a good knowledge of $\Omega_{M}$.

\section{Summary and conclusions}
\label{summary}

We introduce here an Inverse Problem approach to determine $w(z)$ as a
 continuous function in a model--independent and non--parametric way. 
The resulting algorithm retrieves $w(z)$ without imposing any constraints on 
the form of the function. 
The method uses Bayesian information such as the area where this solution 
is to be found, which can be quite unrestrictive. 
The approach explored here enables to see at which z is the maximum information
on $w(z)$ or  $\Delta\Omega_{\Lambda}$ in various samples.

The exploration of dark energy applying this method to the present SNe Ia
sample 
helps to answer the question on whether there is evidence in the evolution 
of $w(z)$ along z. 

 With the 
sets of 162 SNeIa and 192 SNe Ia \cite{davis07, essence} we
 find $w(z)$ compatible at 1$\sigma$ with a cosmological
constant.  The larger data set \cite{davis07} shows improvement in the
resolving kernels at high z.  Error bars in both cases are big enough
to make evident the limited knowledge that 
we have on $w(z)$. However, the bulk
of data to come in the next decade 
and the complementary probes would allow to draw the 
behavior of $w(z)$ and $\Delta\Omega_{\Lambda}(z)$ with much higher
precision. This method can be applied to the determination of $w(z)$ with 
both SNe Ia and BAO measurements.

\ack 
 This work has been supported
 by research grants in cosmology by the Spanish DGYCIT
(ESP20014642--E and BES-2004-4435) and Generalitat de Catalunya 
(UNI/2120/2002). 


\end{document}